\documentclass[10pt]{article}
\usepackage{cite}
\usepackage{epsfig}
\usepackage[dvips,usenames]{color}
\usepackage{color}
\usepackage{rotating}
\usepackage{amsmath,amssymb,amsfonts}
\usepackage{latexsym}
\begin{document}
\begin{center}
 {\Large{\bf {
  Hereditary kernel identification method of
 nonlinear polymeric viscoelastic materials
  }}}
 \end{center}

\vspace{10pt}
\begin{center}
 {\bf Olodo Emmanuel $^{a,1}$,  Villevo  Adanhounme$^{a,2}$, Mahouton Norbert Hounkonnou$^{a,3}$}

\vspace{5pt}
 {\sl\em $^a$International Chair of Mathematical Physics and
Applications},\\ ({\em ICMPA-UNESCO Chair}),
 {\em Universit\'e d'Abomey-Calavi},\\ {\em 072 B.P. 50 Cotonou, Republic of
Benin\\}
{\it E-mail : $^1$olodoe@live.fr; $^2$adanhounm@yahoo.fr;$^3$norbert.hounkonnou@cipma.uac.bj }\newline\newline
\today
\end{center}
\vspace{15pt}
\begin{center}
\begin{abstract}
 This paper deals with a polymeric matrix composite material. The matrix behaviour is described by the modified Rabotnov's nonlinear viscoelastic model assuming the material is  nonlinear viscoelastic. The parameters of creep and stress-relaxation kernels of the model are determined. From the
experimental data related to kernels approximated by spline functions and by means of the method of weighted residual, the formulas for the determination of viscoelastic parameters are derived.
\end{abstract}
\end{center}
 \vspace{10pt}
\pagenumbering{arabic}
\section{Introduction }
Most of polymeric matrix composite materials are characterized by a nonlinear viscoelastic behaviour, even at moderate loading levels. A long term behaviour modeling  these materials requires the determination of viscoelastic characteristics. For the hereditary-type model
  the methods for determining viscoelastic characteristics play a role of great importance and reduce to the establishment and identification of creep and stress-relaxation kernels. The past few years, many authors  studied the nonlinear viscoelastic matrix behaviour modelled as per Schapery's constitutive law. See \cite{1} and references therein. Approaches based on coupling Schapery model and viscoplastic model proposed by Zapas and Crissman are used to predict a long term material behaviour by homogenization\cite{2}.
    In the phenomenon approach of Goldenblat-Kopnov\cite{3},  the hereditary model was developed for  building long term strength phenomenon tests for anisotropic composite materials. This nonlinear model is based on the coupling of plastic potential of Mises-Hill for anisotropic materials and Iliuchin motion approach. In this case, taking into account the peculiarities of material mechanical properties,
     the viscoelastic parameters can be determined on the basis of long term strength tests of material. 

 In a series of  papers \cite{4,5,6,7,8,9}, it was proposed the identification methods for Schapery model using the uniaxial tests of creep-recovery. This approach presents two disadvantages: first, it is too difficult to reproduce exactly the theoretical creep recovery tests,  and
     secondly, one generally deals with the creep recovery tests separately.

 In this paper, in order to avoid these difficulties, we propose a method of hereditary kernel identification for nonlinear viscoelastic materials as follows.
  The experimental data of the kernels are provided by achieving independant tests whose number is equal to the multiple of the number of kernel unknowns. The precision of parameter values of the viscoelastic model requires an efficient method for determining hereditary parameters of nonlinear viscoelastic materials whose matrix behaviour is modelled as per modified Rabotnov's hereditary integral equations with creep and stress-relaxation kernels, taking into account the experimental parameter data for small time values. The experimental data of kernels are approximate by spline functions, and, by means of the method of weighted residual, we obtain formulas for the determination of viscoelastic parameters.

\section{Mathematical model}
For nonlinear viscoelastic materials, the matrix behaviour is modelled by the modified Rabotnov's nonlinear hereditary-type integral equation\cite{10}
   \begin{eqnarray}
\varphi_0(\varepsilon(t))=\sigma(t) + \lambda \int_0^t K(t-\tau)\sigma(\tau)d\tau \label{1}\\
\sigma(t)=\varphi_0(\varepsilon(t)) - \lambda \int_0^t R(t-\tau)\varphi_0(\varepsilon(\tau))d\tau \label{2}
\end{eqnarray}
where $K$ and $ R$ represent the hereditary creep and the stress-relaxation kernels defined,  respectively, by
\begin{eqnarray}
K(t-\tau)= \frac{1}{(t-\tau)^{\alpha}}\sum_{n\geq 0} \frac{(-\beta)^n (t-\tau)^{(1-\alpha)n}}{\Gamma[(1-\alpha)(1+n)]}\\
R(t-\tau)= \frac{1}{(t-\tau)^{\alpha}}\sum_{n\geq 0} \frac{[-(\lambda +\beta)]^n (t-\tau)^{(1-\alpha)n}}{\Gamma[(1-\alpha)(1+n)]}.
\end{eqnarray}
$\lambda$ is a parameter; $\alpha$ and $\beta$ stand for the kernel parameters, $t$ is the time and $\Gamma$ the function defined by
\begin{eqnarray}
\Gamma(z)= \int_0^{\infty} e^{-x}x^{z-1}dx.
\end{eqnarray}
 $\varepsilon$ and $\sigma$ are, respectively, the deformation and the stress-relaxation functions depending on $t$.
 The  scalar function  $\varphi_0,$ depending nonlinearily on $\varepsilon,$ 
is well approximated by a power function  as follows:
  \begin{equation}\label{2a}
\varphi_0(\varepsilon)= \frac{H}{q}\varepsilon^q
\end{equation}
 The coefficients  $H$ and $q$ are    provided by uniaxial traction experimental data.
\section{Hereditary kernel identification}
The choice of efficient method for the kernel parameters determination depends essentially on the physical context.
 So, with respect to stress-relaxation kernel $R$ of model (\ref{1},\ref{2}), 
from 
 the condition   $\varepsilon(t)=\varepsilon(0)=\rm{const},$ 
the time derivative of equation (\ref{2}) gives
\begin{eqnarray}
R(t)=-\frac{1}{\lambda}\frac{q}{H\varepsilon^q}\frac{d\sigma(t)}{dt}\label{3}
\end{eqnarray}
Since the above  expression  of the kernel $R$ is proportional to the stress-relaxation velocity $\frac{d\sigma(t)}{dt}$ with the  coefficient of proportionality remaining constant for $\varepsilon(t)=\rm{const}$, we can determine the kernel parameters of the model  described by (\ref{1}) and (\ref{2}) by the derivation of the experimental curves of the kernel $R$.\\
With respect to the  creep kernel $K$ of this model (\ref{1}) and (\ref{2}),  the physics imposes to  satisfy the condition $\sigma(t)=\sigma(0)=\rm{const}$
that provides  the time derivative of equation (\ref{1}) as follows:
 \begin{eqnarray}
K(t)=\frac{1}{\lambda}\frac{H[\varepsilon(t)]^{q-1}}{\sigma}\frac{d\varepsilon(t)}{dt}\label{4}
\end{eqnarray}
As this expression  of the kernel $K$ is proportional to the creep velocity
$\frac{d\varepsilon(t)}{dt},$ with the coefficient of proportionality remaining variable for $\sigma(t)=\rm{const}$, the above approach is no longer applicable, i. e. we cannot determine the kernel parameters by derivation of the experimental curves of kernel. In this case, the kernel parameters are provided by using the similarity condition between the stress-relaxation isochronic curves and the creep curve of the model  described by (\ref{1}) and (\ref{2}) \cite{10}:
\begin{equation}
\varphi_0(\varepsilon,0)= (1+G(t))\varphi_t(\varepsilon, t) \label{5}
\end{equation}
where $1+G(t)$ is the similarity function defined as
\begin{equation}
1+G(t)= 1 +\lambda\int_0^t K(\tau)d\tau.
\end{equation}
 Here, $\varphi_0(.)$ is the function represented by the deformation curve and $\varphi_t(.),$ the function represented by the creep isochronic curves at every instant $t$.\\

Setting $S(t)= 1+ G(t)$, the condition (\ref{5}) can be differentiated with respect to $t$, yielding the deformation velocity as follows:
\begin{equation}
\frac{d\varepsilon(t)}{dt}= \sigma \bigg\{\frac{d{\varphi_0}^{-1}[S(t)\sigma]}{d[S(t)\sigma]}\bigg\}\frac{dS(t)}{dt},\qquad \sigma=\rm{const}
\end{equation}
 which, taking into account the relation (\ref{4}), allows to write
\begin{equation}
 K(t)= \frac{1}{\lambda}\frac{dS(t)}{dt} \label{6}
\end{equation}

\section{Creep data approximation and method of weighted residual }
In this section, we use spline functions to approximate the creep kernel experimental data. Then, by method of weighted residual, we obtain formulas for the viscoelastic parameter value computation.
 \subsection{Approximation by spline functions }
The similarity of curves representing functions $\varphi_t(.)$ and $\varphi_0(.)$ is considered in the plane $(\varepsilon, \varphi)$ for each fixed  deformation level $\varepsilon_i, i=1\cdots l $
and for the parameter $t_j, j=1,\cdots,n$.
Then data $\varphi_0(\varepsilon_i, 0)=\frac{H}{q}\varepsilon^q_i$ can be approximated by the quantity $\varphi_t(\varepsilon_i, t_j)$. For this purpose we consider the functional  $\Pi$ defined as
\begin{eqnarray}
\Pi\big(\overline{S}_j(t)\big)= \sum_{i=1}^l\big[\varphi_0(\varepsilon_i,0) - \overline{S}_j(t)\varphi_t(\varepsilon_i, t_j) \big]^2,
\end{eqnarray}
where $\overline{S}_j(t)$ is the similarity mean function.
Then the functional minimum is reached for
\begin{equation}
\overline{S}_j(t)= \frac{\sum_{i=1}^l\varphi_0(\varepsilon_i,0)\varphi_t(\varepsilon_i,t_j)}
{\sum_{i=1}^l\big[\varphi_t(\varepsilon_i,t_j)\big]^2}.
\end{equation}
 In order to obtain the best approximation of the discrete data of the similarity mean function, we choose the spline functions defined as\cite{11}
\begin{equation}
\overline{S}_j(t)= \overline{A}_j+  \overline{B}_j(t-t_j)+\overline{C}_j(t-t_j)^2 +\overline{D}_j(t-t_j)^3 \label{7}
\end{equation}
where $\overline{A}_j, \overline{B}_j, \overline{C}_j, \overline{D}_j$ are the coefficients, $j=1,\ldots, n $. By substituting the equation (\ref{7}) into equation (\ref{6}), the experimental data $K(t_j)$ can be estimated
by the functions $ K_j(t,q)$ defined on $[t_j,t_{j+1}]$ as
\begin{equation}
 K_j(t,q)= B_j +2C_j(t-t_j) +3D_j(t-t_j)^2,\,\,\, j=1,\cdots,n  \label{8}
\end{equation}
with the coefficients
\begin{eqnarray}
B_j&=& K(t_j),\\
 2C_j&=& 2\frac{t_j[ K(t_j) -  K(t_{j-1})]}{h_{j-1}\big(2t_j -h_{j-1}\big)},\qquad C_1=0,\\
3D_j&=&\frac{K(t_j) - K(t_{j-1})}{h_{j-1}\big(2t_j - h_{j-1}\big)},\qquad D_1=0, \\
 h_j &=&t_{j+1}-t_j.
\end{eqnarray}
The functions $ K_j(t,q)$ obtained by approximation are presented in Table 1.
 \begin{center}
\begin{tabular}{|r|l|l|l|l|l|l|l|l|l|l|}
\hline
       $j$ & $t_j$& $B_j$ & $2C_j$ & $3D_j$ &$K_j(t,q)$  \\
\hline
1 & 0  & 3750 &  0 & 0 & 3750 \\
\hline
2 & 5 & 3500 &  $-100$ & $ -10$ & $3500- 100(t-5) -10(t-5)^2$ \\
\hline
3 & 7 & 3250 & $-149$ & $-10.42$ & $3250- 149(t-7)-10,42(t-7)^2$  \\
\hline
  4  & 10 & 2900 & $-137$ & $ -6.86$ &$2900-137(t-10) -6.86(t-10)^2$ \\
\hline
5 & 12 & 2600&  $ -167$ & $ -6.82$ & $ 2600-167(t-12) -6.82(t-12)^2$ \\
\hline
6 & 15& 2250& $-130$& $-4.32$ &$2250-130(t-15)-4.32(t-15)^2$\\
\hline
7& 17& 1900 & $-186$ &$-5.47$ & $1900-186(t-17)-5.47(t-17)^2$ \\
\hline
8& 30 & 1500 & $-39.3$ & $-0.65$ & $1500-39.3(t-30)-0.65(t-30)^2$ \\
\hline
9& 70 & 1150& $-12.25$ &$-0.09$ & $1150-12.25(t-70)-0.09(t-70)^2$ \\
\hline
10 & 80& 900 & $-27$& $-0.17$ & $900-27(t-80)-0.17(t-80)^2$ \\
\hline
11 & 100 & 750& $-8.3$ &$ -0.04$ & $750-8.3(t-100)-0.04(t-100)^2$ \\
\hline
12 & 150 & 500& $-6$ &$-0.02$ & $500-6(t-150)-0.02(t-150)^2$ \\
\hline
13 &250 & 300& $-2.5$ & $-0.005$ & $300-2.5(t-250)-0.005(t-250)^2$ \\
\hline
14 & 350 & 250& $-0.6$ &$-0.0008$ & $250-0.6(t-350)-0.0008(t-350)^2$ \\
\hline
15 & 750 & 150 & $-0.34$ & $-0.0002$ & $150-0.34(t-750) -0.0002(t-750)^2$ \\
\hline
16 & 1050& 100 & $-0.2$ &$-0.0001$ &$100-0.2(t-1050)-0.0001(t-1050)^2$ \\
\hline
\end{tabular}

Table 1: Approximated function $K_j(t,q)$
\end{center}
\subsection{Method of weighted residual }
The test of best approximation of data $\{K(t_j),\,\,\,j=1,\ldots,n\}$ by the
functions $\{K_j(t,q),\,\,\,j=1,\ldots,n\}$ remains the method of weighted residual seeking to minimize the residual (error) as
\begin{equation}
 \Omega[\lambda, q]= \sum_{j=1}^n\bigg\{w_j(t)[K(t_j)- \lambda K_j(t,q)]\bigg\}^2,\label{9}
\end{equation}
for a finite set of weighting functions $w_j,\,\,\,j=1,\ldots,n$ defined as
\begin{equation}
  w_j(t)= \bigg\{1+ \bigg|\frac{K(t_j)- \lambda K_j(t,q)}{K(t_{\star})- \lambda K_j(t_{\star},q)}\bigg|^m\bigg\}^{-1} \label{10}
\end{equation}
  satisfying the following conditions
\begin{eqnarray}
  K_j(t,q) \longrightarrow \infty,\qquad  w_j(.)\longrightarrow 0, \\
  K(t_j)= \lambda K_j(t,q),\qquad w_j(t)=1.
\end{eqnarray}
$n$ is the number of discret values of creep kernel for $t$ from $0$ to $t_{\star}=1050$ and $m,$ the order of difference moments, $m=2,3,4,\cdots$.\\
The parameter $\lambda$ and the creep kernel parameter $q$ of the equation (\ref{10}) are defined in two steps:
\begin{itemize}
 \item At the first stage we determine experimentally the initial values $\lambda_0$ and $q_0$ of parameters $\lambda$ and $q$ in the equation (\ref{10}) ;
\item At the second stage, the values $\lambda_0$ et $q_0$ are used, yielding
\begin{equation}
\widetilde{w}_j(t)= \bigg\{1+ \bigg|\frac{K(t_j)- \lambda_0 K_j(t,q_0)}{K(t_{\star})- \lambda_0 K_j(t_{\star},q_0)}\bigg|^m\bigg\}^{-1}. \label{11}
\end{equation}
Then the weighting functions $w_j$ can be provided on the basis of the minimal value $\delta_{\min}$ of residual
\begin{equation}
\delta =\sum_{j=1}^n\bigg\{\widetilde{w}_j(t)[K(t_j)- \lambda_0 K_j(t, q_0)]\bigg\}^2 \label{12}
\end{equation}
 where the sum decreases as $m$ increases.
\end{itemize}
Furthermore, the minimum of residual
\begin{equation}
 \lambda \mapsto \Omega[\lambda,q]= \sum_{j=1}^n\bigg\{\widetilde{w}_j(t)\big[K(t_j)- \lambda K_j(t, q)\big]\bigg\}^2
 \end{equation}
 is reached for
 \begin{equation}
  \widetilde\lambda =\frac{\sum_{j=1}^n{\widetilde{w}_j}^2(t)K(t_j)K_j(t, q) }{\sum_{j=1}^n{\widetilde{w}_j}^
2(t)K^2_j(t, q)}.  \label{13}
  \end{equation}
The very small given value $\gamma$ of residual with given $j$
\begin{equation}
\bigg\{\widetilde{w}_j(t)\big[K(t_j)- \lambda_0 K_j(t,q_0)\big]\bigg\}^2 =\gamma
\end{equation}
 yields
  \begin{equation}
\widetilde{w}_j(t)=\frac{\sqrt{\gamma}}{|K(t_j)- \lambda_0 K_j(t, q_0)|}.\label{14}
 \end{equation}
Taking into account the equations (\ref{13}) and (\ref{14}), we obtain the formula for  the   parameter $\lambda$ value computation
\begin{eqnarray}
\widetilde\lambda &=&\bigg[\sum_{j=1}^n\frac{K(t_j)K_j(t, q)}{|K(t_j)- \lambda_0 K_j(t, q_0)|^2}\bigg]\bigg[\sum_{j=1}^n\frac{ K^2_j(t, q)}{|K(t_j)- \lambda_0 K_j(t, q_0)|^2} \bigg]^{-1},\label{15}\\
t&\in& [t_j,t_{j+1}], j=1,\cdots,n+1.
\end{eqnarray}
For instance, when $t=t_j,$ the formula (\ref{15}) reduces to
\begin{eqnarray}
\widetilde\lambda =\bigg[\sum_{j=1}^n\frac{K^2(t_j)}{|(1 - \lambda_0 )K(t_j)|^2}\bigg]\bigg[\sum_{j=1}^n\frac{ K^2(t_j)}{|(1- \lambda_0)K(t_j)|^2} \bigg]^{-1}=1.
\end{eqnarray}
For each fixed deformation level $\varepsilon_i,\,\,i=1,\cdots,l$ and for the parameter $ t_j,\,\,j=1,\cdots,n$, the equation (\ref{1}) reduces to
\begin{eqnarray}
\frac{H}{q}\varepsilon^q_i&=&\sigma\big[1 + \widetilde\lambda \int_0^{t_j} K_j(\tau,q)d\tau\big] \iff \varepsilon^q_i-\eta_jq=0, \label{16}\\
 \eta_j&=&\frac{\sigma}{H}\big\{1 + \widetilde\lambda [B_jt_j - C_jt^2_j + D_jt^3_j]\big\},\,\,\sigma\mathrm{}=\rm{const}
  \end{eqnarray}
 which admits the solution $\widetilde{q}\in ]0, \overline{q}[$, where  $\overline{q}$ must satisfy the inequality
  \begin{equation}
  \varepsilon^{\overline{q}}_i < \eta_j\overline{q}.
\end{equation}
Therefore we obtain the formula for the parameter $\widetilde{q}$ value computation as follows:
\begin{equation}\label{17}
\varepsilon^{\widetilde{q}}_i-\eta_j\widetilde{q}=0,\,\,\,i=1\cdots,l,\,\,\,j=1,\cdots,16.
\end{equation}
Finally,  we arrive at the following 
 formulas for the computation   of  the parameters $\lambda$ and $q$:
\begin{eqnarray}
\widetilde\lambda &=&\bigg[\sum_{j=1}^n\frac{K(t_j)K_j(t, q)}{|K(t_j)- \lambda_0 K_j(t, q_0)|^2}\bigg]\bigg[\sum_{j=1}^n\frac{ K^2_j(t, q)}{|K(t_j)- \lambda_0 K_j(t, q_0)|^2} \bigg]^{-1},\\
t&\in& [t_j,t_{j+1}], j=1,\cdots,15,\\
\varepsilon^{\widetilde{q}}_i-\eta_j\widetilde{q}&=&0,\,\,\,i=1\cdots,l,\,\,\,j=1,\cdots,16.
\end{eqnarray}
\section{Concluding remarks}
The most important  element  for the kernel identification method of nonlinear viscoelastic models reveals to be  the experimental data $\{K(t_j),\,\,\,j=1,\ldots,n\}$ related to the kernels  which we applied the approximation test to. Among all known standard tests, 
 the cubic spline method gave  the best approximation and using the method of weighted residual to minimize the residual error, we obtained the formulas needed to computate  the parameters $\lambda$ and $q$.

Thus, this study allowed us to  identify the creep and stress-relaxation kernels of the nonlinear viscoelastic materials. Using the spline functions we approximated the experimental data related to the kernel
that permitted, with the help of  the method of weighted residual, to obtain formulas useful for   the parameters computation  of the nonlinear model.


\begin{thebibliography}{00}
\addcontentsline{toc}{section}{References}
\frenchspacing
\small
\addtolength{\itemsep}{-4pt}
\bibitem{1} R.A. Schapery, Nonlinear viscoelastic solids, International Journal of Solids and Structures, 37, Issues1-2,359-366, 2000.
\bibitem{2} A. Pouya and A. Zaoui,  Lin\'earisation et homog\'en\'eisation en visco\'elasticit\'e, Comptes Rendus de l'Acad\'emie des Sciences, t.326 (Serie IIb): 365-370, 1999.
    \bibitem{3} V.A. Kopnov,  Long term strength of anisotropic materials with a complex stress state. Strength of materials, 14, N�2, pp.183-187, 1982.
\bibitem{4} R.M. Haj-Ali and A.H. Mulina, Numerical finite element formulation
of Schapery nonlinear viscoelastic material model, International Journal of Numerical Methods in Engineering , 59: 25-45, 2004.
\bibitem{5} J. Lai and A. Bakker,
 An integral constitutive equation for nonlinear plasto-viscoelastic behavior of High-Density polyethylene.
 Polymer Engineering and Science, 33(17): 1339-1347, 1995.
\bibitem{6} J. Lai and A. Bakker,
3-D Schapery representation for nonlinear viscoelasticity and finite element implementation, Computational Mechanics, 18: 182-191, 1996.
   \bibitem{7} R. Mohan and D.F. Adams, Nonlinear creep-recovery response
     of a polymer matrix and its composites. Experimental Mechanics, 25(4): 262-271,1985.
  \bibitem{8} S.P.Zaoutsos, G.C. Papanicolaou, and A.H. Cardon, On the nonlinear viscoelastic characterization of polymer matrix composites.
         Composites Science and Technology, 58(6): 883-889, 1998.
\bibitem{9} L.J.Zapas and J.M. Crissman,  Creep and recovery behavior of ultra-high
molecular weight polyethylene in the region of small uniaxial deformations, Polymer, 25: 57-62,1983.
\bibitem{10} Yu V.  Suvorova, Yu N. Rabotnov's nonlinear hereditary-type equation
and its applications, Mechanics of Solids, 39(1): 132-138, 2004.
\bibitem{11} V. Parton and P. Perline,  Equations int\'egrables de la th\'eorie de l'\'elasticit\'e. Editions Mir, Moscou, 1977.
 \bibitem{12} R. Brenner, O. Castelneau, and P. Gilormini, 
  A modified affine theory for the overall properties of nonlinear
  composites. Comptes rendus de l'Acad�mie des Sciences, t.329
  (S�rie II b): 649-654, 2001.
  \bibitem{13} C.P. Buckley,
  Multiaxial nonlinear viscoelasticity in solid polymers.
  Polymer Engineering and Science, 27(2): 155-164, 1987.
   \bibitem{14} R.M. Christensen, Mechanics of composite materials.
   John Wiley and sons, New-York, 1980.
  \bibitem{15}  A.D. Drozdov,  A constitutive model for nonlinear
    for viscoelastic media.
    International Journal of Solids and Structures, 34(21): 2685-2707, 1997.
    \bibitem{16}A.D. Drozdov, A. Al-Mulla, and R.K. Gupta,  A constitutive model for the viscoplastic behaviour of rubbery polymers at finite strains. Acta Mechanica, 164: 139-160, 2003.
  \bibitem{17} S. Lee and W.G. Knauss,  A note of the determination of relaxation and creep data from ramp tests.
      Mechanics of Time-Dependent
Materials, 4:1-7, 2000.

 \bibitem{18} F.J. Lockett, Creep and stress-relaxation experiments for nonlinear materials. International Journal of Engineering Science, 3:59-75, 1965.

\end{thebibliography}
\end{document}